\documentclass[10pt,fleqn]{article}

\usepackage{latexsym}
\usepackage{amsbsy}
\usepackage{amsfonts}
\usepackage{amssymb}
\usepackage{amscd}
\usepackage{mathabx}  
\usepackage{t1enc}
\usepackage{epsfig}
\usepackage{color}
\usepackage{cite}

\usepackage{titling}

\usepackage{capt-of}
\usepackage{caption}

\def\appendix{\par
  \setcounter{section}{0}%
  \def\@chapapp{\appendixname}%
\def\thesection{Appendix \Alph{section}}}

\renewcommand{\appendixname}{Appendix}

\newcommand{\greeksym}[1]{{\usefont{U}{psy}{m}{n}#1}}

\newcommand{\umuppf}{\mbox{\small\greeksym{m}}}
\def\micron{{\umuppf}{\rm m}}

\mathchardef\Gamma="0100
\mathchardef\Delta="0101
\mathchardef\Theta="0102
\mathchardef\Lambda="0103
\mathchardef\Xi="0104
\mathchardef\Pi="0105
\mathchardef\Sigma="0106
\mathchardef\Upsilon="0107
\mathchardef\Phi="0108
\mathchardef\Psi="0109
\mathchardef\Omega="010A

\newcommand{\sq}{\hbox{\rlap{$\sqcap$}$\sqcup$}}
\newcommand{\qed}{\ifmmode\sq\else{\unskip\nobreak\hfil
  \penalty50\hskip1em\null\nobreak\hfil\sq
  \parfillskip=0pt\finalhyphendemerits=0\endgraf}\fi{}}

\def\I{{\rm i}}

\def\D{{\rm d}}
\def\E{{\rm e}}

\def\vec{\boldsymbol}

\def\cal{\mathcal}

\def\its{\it}

\def\qed{\hfill $\Box$}

\def\primespe{\kern .08em '}

{\bf}{\it}
{\bf}{\it}

\newtheorem{remark}{Remark}[section]

\input{diagxy}

\begin{document}

\title{Graphical Constructions of Wavefronts and Waist Parameters in Gaussian Beam Optics}

\date{}
\maketitle

\begin{center}

\vskip -2.4cm

{
\renewcommand{\thefootnote}{}
 {\bf   Pierre Pellat-Finet\footnote{\hskip -.53cm Laboratoire de Math\'ematiques de Bretagne Atlantique (LMBA) UMR CNRS 6205,

\noindent Universit\'e de Bretagne Sud, CS 60573, 56017 Vannes, France.

\noindent pierre.pellat-finet@univ-ubs.fr     
 }}
}
\setcounter{footnote}{0}

\medskip
    {\sl \small Universit\'e  Bretagne Sud,    UMR CNRS 6205, LMBA, F-56000 Vannes, France}

\smallskip
\end{center}


\begin{center}
\begin{minipage}{12cm}
\hrulefill

\smallskip
{\small
  {\bf Abstract.}   We provide several diagrams for the graphical determination of certain ele\-ments of a Gaussian beam based on prior knowledge of other elements. For example, these diagrams allow us to determine the plane of the beam waist and the Rayleigh range from knowledge of two wavefronts constituting the beam, or to determine the size of the light spot on a given wavefront. We also present a simple method for determining the waist position and the Rayleigh range of the image of a Gaussian beam formed by a lens.

\smallskip
\noindent {\bf Keywords.}  Beam waist, double conjugation law, Fourier optics, Gaussian beam, Rayleigh range. 

}
\hrulefill
\end{minipage}
\end{center}

\section{Introduction}

Mathematical formulas describing Gaussian beam properties are well known and allow one to precisely evaluate characteristic parameters
of a given Gaussian beam. However, there is interest in developing graphical methods to determine certain elements of a Gaussian beam in a direct and easily graspable way. For example, this includes determining the location of the beam waist, based on the knowledge of two wavefronts, or alternatively the wavefront at a given distance from  the waist plane.

The determination of the waist of an optical resonator from the knowledge of the relative positions of the two mirrors constituting the resonator can be achieved by the ``circle method,'' which is reported by 
Siegman \cite{Sie} and   Metcalf {\its et al.} \cite{Met} (the method is implicitly applied to a Fabry-Perot cavity  by Laures \cite{Lau}). 
Their methods are based on usual formulas for optical resonators and can be directly applied to Gaussian beams. A graphical determination of the waist of the image beam formed by a lens is described by Arnaud \cite{Arn}. It is grounded in imaging  skew rays (or ``complex'' rays) through a lens. Constructions based on geometrical optics have been developed by Laures \cite{Lau}.

We will provide several examples of such graphical constructions.  We complement the result of Metcalf {\its et al.} \cite{Met}  with the graphical construction of the size of the light spot  (more precisely: the reduced transverse radius) on a given wavefront. We also describe a simple  graphical construction of the waist position and the Rayleigh range of the image of a given  Gaussian beam formed by a lens. The method we use is derived from usual formulas for Gaussian beams, but takes advantage of the double conjugation law that characterizes coherent geometrical imaging \cite{PPF2,PPF3,PPF1}.

\section{Gaussian beams}\label{sect2}

We consider Gaussian beams that are produced by lasers with stable resonators. A simple way to determine whether  a resonator made up of two spherical mirrors ${\cal M}_1$ (vertex $V_1$, center of curvature $C_1$) and ${\cal M}_2$ (vertex $V_2$, center of curvature $C_2$) is stable is as follows \cite{PPF3,PPF4}: points $V_j$ and $C_j$ ($j=1,2$) are arranged along the optical axis so that the subscripts are ordered as 1212 or 2121. For example, the arrangements $V_1V_2C_1C_2$ or $C_2V_1V_2C_1$ correspond to stable resonators, whereas the arrangement $V_1V_2C_2C_1$ corresponds to an unstable resonator.

The mirrors, which necessarily lie at a certain distance from each other ($V_1\ne V_2$),
are concentric if $C_1=C_2$. They are confocal if $V_1=C_2$ and $V_2=C_1$; the reason is that they share their focus, since the focus  of a mirror is the midpoint between the vertex and the center of curvature. Concentric and confocal mirror resonators constitute limiting cases of stability and should generally be treated separately (see Section \ref{sect6}).

A Gaussian beam can be characterized by a sequence of wavefronts that are non-concentric spherical caps (this finding is valid in the framework of a metaxial theory of diffraction \cite{PPF1,PPF2,PPF3}, i.e. a second order aproximation with respect to transverse magnitudes). 
The mirror surfaces of a laser are wavefronts of the Gaussian beams it produces. By comparison, two wavefronts are said to be confocal if the vertex of one coincides with  the center of the other.

For the ``fundamental beam''  (corresponding to the fundamental mode of the stable optical re\-so\-nator generating the beam), if $z$ denotes the direction of propagation,  the field  amplitude on the (spherical)  wavefront at abscissa $z$ takes the form
\begin{equation}
  U_z(\vec r)=U_0\exp\left(-{r^2\over w^{2}}+\I \alpha\right)\,,\label{eq0}
\end{equation}
where a factor $\exp (-\I k z)$ is omitted ($k$ denotes the wave number), $\alpha$ is known as the Gouy phase (which depends on $z$), and $U_0$ is a dimensional constant. In Eq. (\ref{eq0}), $\vec r$ is a transverse vector, related to Cartesian coordinates: $\vec r=(x,y)$ and $r=\|\vec r\|=(x^2+y^2)^{1/2}$. The parameter $w$ depends on $z$  and represents the transverse radius of the light spot at abscissa $z$ (this radius is  defined as the distance  at which  the amplitude modulus drops to $1/{\rm e}$ of its value at the center: $|U_z(w)|= U_0/\E$ ; the irradiance drops to $1/{\rm e}^2$).

\begin{figure}[t]
  \centering
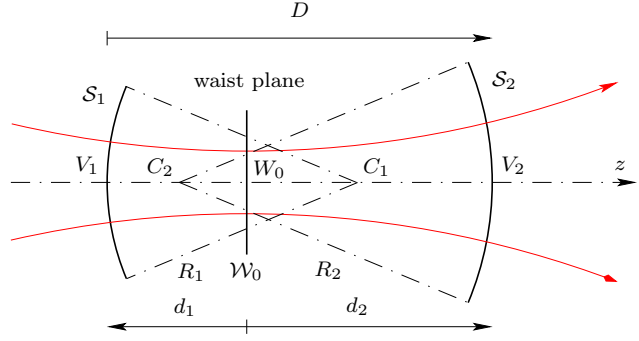
\caption{\small A Gaussian beam. Spherical caps ${\cal S}_1$ and ${\cal S}_2$ are wavefronts. The wavefront ${\cal W}_0$, located at $W_0$, is a plane and contains the waist. Algebraic measures are positive if taken in the sense of light propagation. In the diagram, light is assumed to propagate from left to right, so that  $D=\overline{V_1V_2}>0$, $d_1=\overline{W_0V_1}<0$ and $d_2=\overline{W_0V_2}>0$. Radii of curvature of wavefronts are $R_1=\overline{V_1C_1}>0$ and $R_2=\overline{V_2C_2}<0$.
  \label{fig1}}
\end{figure}

\medskip
\noindent {\bf Graphical convention}: {\its All Gaussian beams considered in this article have rotational symmetry about the optical axis $z$. All subsequent diagrams are drawn in a meridional plane (i.e. a plane that contains the axis of rotational symmetry). In such diagrams, spherical caps are represented by circular segments, and spheres are represented by circles. For example, in Fig.\  \ref{fig1}, circular segments labeled as ${\cal S}_1$ and ${\cal S}_2$ are merely the traces  of spherical caps ${\cal S}_1$ and ${\cal S}_2$ in a meridional plane. What is indicated as the ``waist plane'' is  the trace of the waist plane in  a
  meridional plane.}

\medskip

A basic concept for Gaussian beams (and stable resonators) is the waist. The waist is where the beam is narrowest.  It can be proved that the waist lies on a plane, which is a wavefront.
We denote by ${\cal W}_0$ the (plane) wavefront at the waist and by $w_0$ the transverse radius of the waist. The waist intersects the axis at $W_0$ (Fig.\ \ref{fig1}).
The field amplitude in the waist plane corresponding to the fundamental mode is
\begin{equation}
  U(\vec r)=U_0\exp\left(-{r^2\over w_0^{\; 2}}\right)\,,\label{eq1}
  \end{equation}
if the plane of the waist is taken as origin for the phase (Gouy phase).

A useful parameter is the Rayleigh range (or Rayleigh length) $\zeta_0$ (also denoted as $z_R$ in many texts) such that
\begin{equation}
  \zeta_0={\pi\over \lambda}w_0^{\; 2}\,,\end{equation}
where $\lambda$ is the wavelength of the radiation in the propagation medium.

The confocal parameter is $2\zeta_0$; it is the distance between two confocal wavefronts that are symmetric  with respect to the waist plane (see Section \ref{sect43}).

In general $\zeta_0$ is much more larger than $w_0$. For example, at $\lambda = 1\, \micron$ and for $w_0=1\,{\rm mm}$, we obtain $\zeta_0\approx 3.14\,{\rm m}$. The confocal parameter is $2\zeta_0\approx 6.28\,{\rm m}$.

Wavefronts of a given Gaussian beam are spherical caps (in the framework of a metaxial approximation, as mentioned above). The radius of curvature of the wavefront ${\cal S}_d$ at a distance $d$ from the waist plane is
\goodbreak
\begin{equation}
  R_d=-d-{\pi^2w_0^{\; 4}\over \lambda^2d}=-d -{\zeta_0^{\; 2}\over d}\,.\label{eq4}
\end{equation}
In fact  $d$ is an algebraic distance (or algebraic measure), taken from the waist plane to the vertex of ${\cal S}_d$. It is positive if orientated in the direction of light propagation, or negative in the opposite direction.

\medskip
\noindent {\bf Rule}: {\its Algebraic distances (or algebraic measures) along the optical axis are positive if taken in the sense of light propagation.}

\medskip
Very often, by  abuse of language, we write  ``distance'' instead of  ``algebraic distance''.
The previous rule  holds for algebraic distances between two points on the optical axis. It also holds for the radii of curvature of spherical caps, taken from vertices to centers of curvature (assumed to lie on the optical axis).

Formulas for optical resonators and Gaussian beams can be found in many articles or textbooks \cite{Sie,PPF3,Ana,Kog2,Ger}. They may  differ slightly because algebraic sign conventions vary from one author to another. In general, a Gaussian beam can be thought of as being generated by a laser resonator whose mirror surfaces  coincide  with two arbitrary  non-confocal wavefronts of the beam. In this way, formulas established for the mirrors  of a resonator also apply to the  wavefronts of the corresponding Gaussian beam, and vice versa. For example, in Eq.\ (\ref{eq4}), $d$ may be interpreted as the algebraic distance  from the waist plane to one mirror of the resonator whose radius of curvature is $R_d$.

We finally mention the following finding, which will be implicitly used:
let ${\cal S}_1$ and ${\cal S}_2$ be two circular segments that are the traces of two non-confocal wavefronts of a Gaussian beam in a meridional plane  (see Fig.\ \ref{fig1}). Then the circles  ${\cal C}_1$ and ${\cal C}_2$  with respective diameters $V_1C_1$ and $V_2C_2$ intersect at two points (an example is given in Fig.\ \ref{fig2}).

A proof is as follows. For $j=1,2$, Eq.\ (\ref{fig1}) gives
\begin{equation}
  R_j=-d_j-{\pi^2w_0^{\; 4}\over \lambda^2d_j}=-d_j -{\zeta_0^{\; 2}\over d_j}\,.
\end{equation}
so that
\begin{equation}
  \overline{W_0C_j}=d_j+R_j=-{\zeta_0\over d_j}\,.\end{equation}
Let us assume that $\overline{W_0V_1}=d_1<0$ and $\overline{W_0V_2}=d_2>0$ (as in Fig.\ \ref{fig1}). Then $\overline{W_0C_1}>0$ and $\overline{W_0C_2}<0$. We conclude that circles ${\cal C}_1$ and ${\cal C}_2$ intersect at two points $I$ and $J$ (see Fig.\ \ref{fig2}).

Next, we assume $d_1<d_2<0$. Then  $\overline{W_0C_2}>\overline{W_0C_1}>0$.  We conclude that circles ${\cal C}_1$ and ${\cal C}_2$ intersect at two points. The result is valid if $d_2<d_1<0$. The same method shows that circles ${\cal C}_1$ and ${\cal C}_2$ intersect at two points if $d_1>d_2>0$, or if $d_2>d_1>0$.

It can be shown that ${\cal C}_1$ and ${\cal C}_2$ intersect at two points if, and only if, points $V_j$ and $C_j$ ($j=1,2$) are arranged along the optical axis as for a stable resonator (i.e. their subscripts are ordered as 1212 or 2121, as mentioned earlier in this section).  Since a Gaussian beam is generated by a stable laser resonator, the previous circles associated to it always intersect.

\section{Determining the waist position and the Rayleigh range from the knowledge of two wavefronts}\label{sect3}

\subsection{Formulas for the location of the waist}\label{sect31}

Let ${\cal S}_1$ and ${\cal S}_2$ be two wavefronts of a Gaussian beam whose waist plane ${\cal W}_0$ is at $W_0$ (Fig.\ \ref{fig1}). The vertex of ${\cal S}_j$ is $V_j$, its center of curvature is $C_j$, and its radius of curvature is $R_j=\overline{V_jC_j}$.  Let $D=\overline{V_1V_2}$.
Let $d_1$ be the algebraic distance from ${\cal W}_0$ to ${\cal S}_1$. Then \cite{PPF2,PPF3}
\begin{equation}
  d_1=-{D(D+R_2)\over 2D-R_1+R_2}\,.\label{eq7}
  \end{equation}
The algebraic distance from ${\cal W}_0$ to ${\cal S}_2$  is $d_2$ such that $D=d_2-d_1$, that is
\begin{equation}
  d_2=d_1+D={D^2-DR_1\over 2D-R_1+R_2}={D(D-R_1)\over 2D-R_1+R_2}\,.\label{eq8}
\end{equation}

\subsection{Determination of the waist location from two wavefronts}\label{sect32}
Let ${\cal S}_1$ and ${\cal S}_2$ be two non-confocal wavefronts of a Gaussian beam as before (Figs.\ \ref{fig1} and \ref{fig2}).
Let ${\cal C}_j$ be the circle of diameter $V_JC_j$. As explained in the previous section, circles ${\cal C}_1$ and ${\cal C}_2$ intersect at points $I$ and $J$ (see Fig.\ \ref{fig2}).  Let $K$ be the intersection of $IJ$ with the axis $z$. We denote    $q_1=\overline{KV_1}$ and  $q_2=\overline{KV_2}$.
Then
\begin{equation}\overline{KC_1}=\overline{KV_2}+\overline{V_2V_1}+\overline{V_1C_1}=q_2-D+R_1\,,
\end{equation}
\begin{equation}
\overline{KC_2}=\overline{KV_1}+\overline{V_1V_2}+\overline{V_2C_2}=q_1+D+R_2\,.
\end{equation}
We observe that  $K$ necessarily lies between $V_1$ and $C_1$, so that the power of point $K$ with respect to the circle ${\cal C}_1$ is  $\overline{KV_1}\cdot\overline{KC_1}=-KI\cdot KJ$. For the same reason, its power with respect to ${\cal C}_2$ is $\overline{KV_2}\cdot\overline{KC_2}= -KI\cdot KJ$. Therefore $\overline{KV_1}\cdot\overline{KC_1}=\overline{KV_2}\cdot\overline{KC_2}$, which yields
\begin{equation}
q_1(q_2-D+R_1)=-KI\cdot KJ=q_2(q_1+D+R_2)\,,\label{eq11}
\end{equation}
and then $q_1(R_1-D)=q_2(D+R_2)$, that is,
\begin{equation}
  {q_1\over q_2}={D+R_2\over R_1-D}\,.\label{eq9}
    \end{equation}
Since $D=q_2-q_1$, Eq.\ (\ref{eq9}) leads to
\begin{equation}
  q_1=-{D(D+R_2)\over 2D-R_1+R_2}\,,\hskip .5cm {\rm and}\hskip .5cm
q_2={D(D-R_1)\over 2D-R_1+R_2}\,.
  \end{equation}
 According to  Eqs.\ (\ref{eq7}) and (\ref{eq8}), we conclude that $q_1=d_1$ and $q_2=d_2$, which means that the plane of the beam waist is located at $K$. 

\subsection{Rayleigh range}

The Rayleigh range $\zeta_0$ ($\zeta_0>0$) is given by \cite{PPF3}
\begin{equation}
  \zeta_0^{\; 2}={D(R_1-D)(D+R_2)(D-R_1+R_2)\over (2D-R_1+R_2)^2}\,.
\end{equation}

We use the notation introduced in  Sections \ref{sect31} and \ref{sect32}. Since $D=d_2-d_1$, Eq.\ (\ref{eq7}) yields
\begin{equation}
  d_2-D+R_1=d_1+R_1=-{D(D+R_2)\over 2D-R_1+R_2}+R_1= {(R_1-D)(D-R_1+R_2)\over   2D-R_1+R_2}\,.
\end{equation}
Since $q_1=d_1$ and $q_2=d_2$, we use Eqs.\ (\ref{eq7}) and  (\ref{eq11}) to obtain
\begin{equation}
  KI^2=KI\cdot KJ=-d_1(d_2-D+R_1)= {D(D+R_2)\over  2D-R_1+R_2}\,{(R_1-D)(D-R_1+R_2)\over   2D-R_1+R_2}
  =\zeta_0^{\; 2}\,.
\end{equation}
Both $\zeta_0$ and $KI$ are positive, so that $\zeta_0=KI$.

\subsection{Graphical construction}

\begin{figure}[h]
  \centering
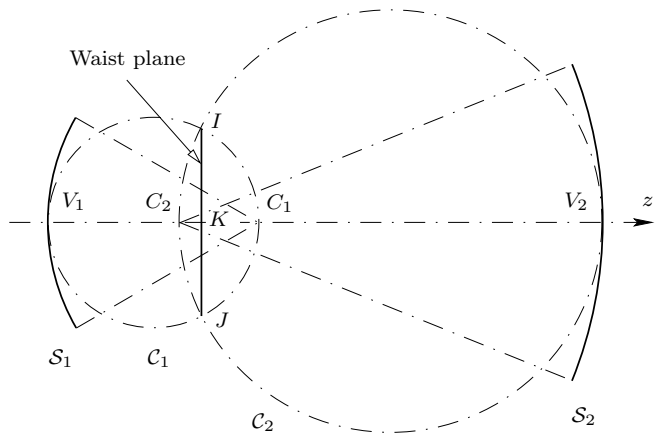
\caption{\small If ${\cal S}_1$ and ${\cal S}_2$ are the traces of two wavefronts of a Gaussian beam, the trace of the waist plane  is the straight line  passing through the points $I$ and $J$,  defined as the intersections of the circles ${\cal C}_1$ and ${\cal C}_2$ having diameters $V_1C_1 $ and $V_2C_2 $. The waist is located at $W_0\equiv K$. The length of segment $KI$ is equal to the Rayleigh range $\zeta_0$. (The confocal parameter is $2\zeta_0=IJ$.) The diagram above is drawn for $R_2<0<R_1$. However, the method is also valid if $R_1$ and $R_2$ are of the same sign. 
  \label{fig2}}
\vskip -.4cm
\end{figure}

The corresponding constuction for the waist location and the Rayleigh range  is given in Fig.\ \ref{fig2}.
We consider two  non-confocal  wavefronts ${\cal S}_1$ and ${\cal S}_2$ of a Gaussian beam, located at $V_1$ and $V_2$, with centers of curvature $C_1$ and $C_2$. The drawings are as follows (this is the ``circle method'' \cite{Sie}):
\begin{enumerate}
\item We draw the circle ${\cal C}_1$ whose diameter is $V_1C_1$, and the circle ${\cal C}_2$ whose diameter is $V_2C_2$.
\item Circles ${\cal C}_1$ and ${\cal C}_2$ intersect at points $I$ and $J$.
\item The plane of the waist is located at $K$, the point where $IJ$ intersects the axis.  This plane is  orthogonal to the axis at $K$. Then $W_0\equiv K$.
   \item The Rayleigh range is the length of $KI$. (We also have $IJ=2IK=2\zeta_0$.)
  \end{enumerate}
 
\begin{remark}\label{rem31} {\rm Let us consider a stable   optical resonator with non-confocal mirrors ${\cal M}_1$ and ${\cal M}_2$. The surfaces of the mirrors are wavefronts for Gaussian modes propagating in the resonator. The circle method described above can be applied with ${\cal S}_1\equiv{\cal M}_1$ and ${\cal S}_2={\cal M_2}$ to determine the position of the  waist and the Rayleigh range of the resonator \cite{Sie,Met}.}  \end{remark}

\begin{remark}\label{rem32} {\rm In general, two spherical caps (with a common axis $z$) determine a unique waist plane and a unique Rayleigh range, which do not depend on the wavelength, according to the construction of Fig.\ \ref{fig2}.

    On the other hand, the waist radius $w_0$ is deduced from the Rayleigh range $\zeta_0$ according to
    \begin{equation}
      w_0=\sqrt{\lambda \zeta_0\over \pi},
    \end{equation}
    and depends on the wavelength in the propagation medium. This means that given two sphe\-ri\-cal caps, there are many Gaussian beams that admit these surfaces as wavefronts (there is one beam for each wavelength), and all the corresponding waists lie in a single plane.
}
\end{remark}

\begin{remark}\label{rem33}{\rm If ${\cal S}_1$ and ${\cal S}_2$ are confocal, circles ${\cal C}_1$ and ${\cal C}_2$ coincide, causing the construction method developed in this section to fail to locate the waist plane. Nevertheless, if the waist plane is known, the Rayleigh range can be drawn.
} \end{remark}

\section{Determining a wavefront from the knowledge of the waist plane and the Rayleigh range}\label{sect4}

\subsection{Wavefront}\label{sect41}

\begin{figure}[h]
  \centering
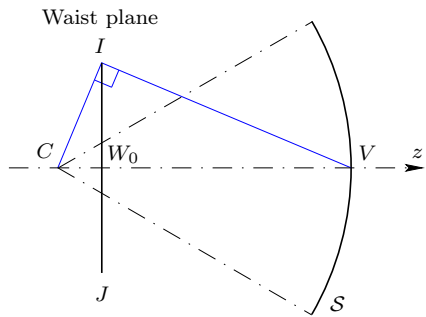
\caption{\small Construction of a wavefront ${\cal S}$ knowing the waist plane (located at $W_0$) and the Rayleigh range $\zeta_0$. The scale is given by   $W_0I=W_0J=\zeta_0$.\label{fig3}}
\end{figure}

In Fig.\ \ref{fig2}, triangles $V_1IC_1$ and $V_2IC_2$ are right triangles, because  $V_1C_1$ and $V_2C_2$ are diameters of circles ${\cal C}_1$ and ${\cal C}_2$.  That property leads us to the subsequent constructions, where  the Rayleigh range  $\zeta_0$ is given beforehand and $W_0I=IJ/2=\zeta_0$  (see Fig.\ \ref{fig3}):
\begin{itemize}
\item Construction of the wavefront centered at a given point $C$ (on the axis):
 \begin{enumerate}
  \item Draw the segment $CI$.
  \item The point $V$ is the intersection of the axis $z$ with the straight line perpendicular to $CI$ at $I$.
  \item The wavefront $S$ is centered at $C$ and its vertex is $V$.
 \end{enumerate}
\item Construction of the wavefront with vertex at $V$ (on the axis):
  \begin{enumerate}
  \item Draw the segment $VI$.
  \item The point $C$ is the point where the straight line perpendicular to $VI$ at $I$ intersects the axis. 
  \item The wavefront $S$ is centered at $C$ and its vertex is $V$.
 \end{enumerate}
 \end{itemize}

\subsection{Reduced transverse radius}

The transverse beam radius on the wavefront ${\cal S}$ at a distance $d$ from the waist is $w_d$ such that \cite{Sie,Ana,PPF3,Ger}
\begin{equation}
  w_d^{\; 2}=w_0^{\; 2}+{\lambda^2 d^2\over \pi^2w_0^{\; 2}}\,,\label{eq19}\end{equation}
so that the reduced transverse radius is defined by
\begin{equation}
  \zeta_d={\pi\over \lambda} w_d^{\; 2}=\zeta_0+{d^2\over \zeta_0}\,.\label{eq20}
\end{equation}
The reduced transverse radius $\zeta_d$ generalizes the Rayleigh range $\zeta_0$ to  an arbitrary wavefront. 
It  provides the dimension of the light spot on the wavefront. It is independant of the wavelength.

Equation (\ref{eq20}) yields
\begin{equation}
  \zeta_0\zeta_d =\zeta_0^{\; 2}+d^2\,.
\end{equation}
The waist is located at $W_0$ and  $|d|=W_0V$ (Fig.\ \ref{fig3}).  Since  $W_0I=\zeta_0$, and since $IW_0V$ is a right triangle, we deduce $VI^2=\zeta_0^{\; 2}+d^2=\zeta_0\zeta_d$.

The issue is then to construct a segment whose length is $\zeta_d$ and such that $\zeta_0\zeta_d=VI^2$, where $\zeta_0$ and $VI$ are known. We observe that $VI^2$ is the power of $I$ with respect to every circle tangent to $VI$ at $V$. Since $W_0I=\zeta_0$, we conclude that $\zeta_0\zeta_d$ is the power of $I$ with respect to the circle ${\cal C}$ tangent to $VI$ at $V$ that passes through $W_0$. The construction of $\zeta_d$ is as follows (see Fig.\ \ref{fig4}):

\begin{figure}[h]
      \begin{minipage}{8cm}
        \centering
            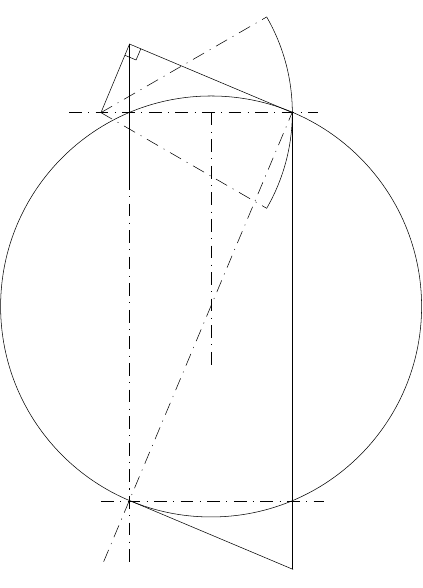
            \caption{\small Explanation for the construction of the reduced transverse radius $\zeta_d$ on the wavefront $S$ located at a distance $d$ from the waist plane ($d=\overline{W_0V}$). We obtain $\zeta_d=VJ'$.\label{fig4}}
      \end{minipage}
      \hfill
      \begin{minipage}{6cm}
        \vskip .2cm
         \centering
          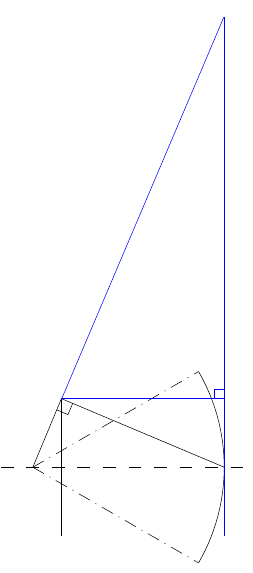
          \caption{\small The reduced transverse radius on ${\cal S}$ is $\zeta_d=VI'$. The point $I'$ is the intersection of the straight lines supporting the segments $CI$ and $VP$.\label{fig5}}
            \end{minipage}
  \end{figure}

\begin{enumerate}
\item Construction of the circle ${\cal C}$:
  \begin{itemize}
  \item Draw the straight line $\Delta$, perpendicular to $VI$ at $V$.
  \item Draw the straight line $\Delta '$, perpendicular to $W_0V$ at $M$, where $M$ is the midpoint of segment $W_0V$.
  \item The center $O$ of ${\cal C}$  is the intersection of $\Delta$ and $\Delta '$.
    \item Draw the circle ${\cal C}$ whose center is $O$ and radius is $O V=O W_0$.
  \end{itemize}
  \item The straightline $\Delta ''$ that extends segment  $W_0I$ intersects ${\cal C}$ at $W_0$ and $L$. The power of $I$ with respect to ${\cal C}$ is $IW_0\cdot IL=VI^2=\zeta_0\zeta_d$, and since $IW_0=\zeta_0$, we obtain $IL=\zeta_d$.
\item The straight line orthogonal to  $\Delta$ at $L$ (which is tangent to circle ${\cal C}$) intersects $VN$ at $J'$ and $VJ'=\zeta_d$.
\end{enumerate}

Finally, the construction of the reduced transverse radius on ${\cal S}$ is very simple  (see Fig.\ \ref{fig5}):  the straight line 
that extends the segment $CI$ intersects the straight line tangent to ${\cal S}$ at $V$ at $I'$, and $VI'=\zeta_d$.   This can be understood by comparing Figs.\ \ref{fig4} and \ref{fig5}:
\begin{enumerate}
\item Let $P$ be the orthogonal projection of $I$ on the straight line tangent to the wavefront ${\cal S}$ at $V$. Then $IP=W_0V$
\item Since $\angle VLN=\angle ICV=\angle I'IP$, right triangles $VLN$ (Fig.\ \ref{fig4}) and $I'IP$ (Fig.\ \ref{fig5}) are equal, so that $I'P=VN$.
  \item Since $NJ'=IW_0=PV$, we obtain $\zeta_d=VJ'=I'V$.
\end{enumerate}

\subsection{Rayleigh spheres}\label{sect43}
\begin{figure}[h]
  \centering
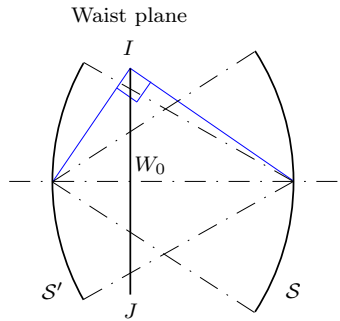
\caption{\small Spherical caps ${\cal S}$ and ${\cal S}'$ are symmetric confocal spheres. The reason is that if these spherical caps were mirrors, they would share a common focus, as the focus of a spherical mirror is located at the midpoint between the vertex and the center. These spherical caps are wavefronts of a Gaussian beam whose waist is located at $W_0$ and Rayleigh range is $\zeta_0=W_0I$.\label{fig6}}
\end{figure}

The constructions given in Section \ref{sect41} (see Fig.\ \ref{fig3}) show that,  exchanging the vertex $V$ with the center of curvature $C$ of a given wavefront ${\cal S}$ provides another wavefront ${\cal S}'$ of the considered Gaussian beam (Fig.\ \ref{fig6}). The radii of curvature of ${\cal S}$ and ${\cal S}'$ are opposite and the two  spherical caps are confocal.  According to the metaxial theory of Fourier optics, the field amplitudes on these  caps are related by an optical Fourier transformation \cite{PPF2,PPF3}. 

\begin{figure}[h]
  \centering
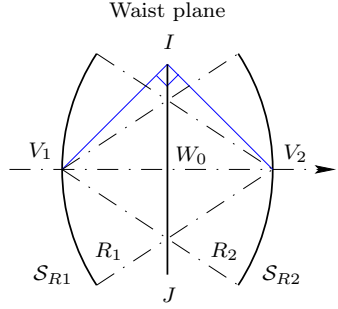
\caption{\small Rayleigh spheres ${\cal S}_{R1}$ and ${\cal S}_{R2}$  are  confocal spheres such that $W_0V_1=W_0V_2=W_0I=\zeta_0$. Then $\angle V_1IW_0=\angle W_0IV_2=\pi /4$. The radii of curvature are such that $R_1=-R_2=2\zeta_0$. \label{fig7}}
\end{figure}

Two confocal wavefronts deserve to be mentioned: we call them ``Rayleigh spheres''. The absolute values of their radii of curvature are equal to the confocal parameter, that is, twice the Rayeigh range $\zeta_0$. It can be shown that among all the wavefronts of a Gaussian beam, the smallest absolute value of their radii of curvature corresponds to the Rayleigh spheres. The proof is deduced from Eq.\ (\ref{eq4}) as follows. The derivative of $R_d$ with respect to $d$ is \goodbreak 
\begin{equation}
  {\D R_d\over\D d}=-1+{\zeta_0^{\;2}\over d^2}\,,\end{equation}
and is zero for $d=\pm \zeta_0$. For $d=d_1=-\zeta_0$, we obtain $R_{1}=2\zeta_0 >0$, and for $d=d_2=\zeta_0$, we obtain $R_{2}=-2\zeta_0 <0$ (see Fig.\ \ref{fig7}). We have $\overline{V_1V_2}=-d_1+d_2=2\zeta_0$.
According to the construction of Fig.\ \ref{fig5}, we obtain that the reduced transverse radius on a Rayleigh sphere is $\zeta_R=2\zeta_0$, so that the corresponding transverse radius is $w_R=\sqrt{2}\,w_0$.

\begin{remark} {\rm The prior knowledge of the waist plane and the Rayleigh range enables the construction of the Rayleigh spheres. However,   according to Remark \ref{rem33}, knowledge of two confocal wavefronts alone is insufficient to determine the waist plane unless it is known that these confocal spheres are the Rayleigh spheres (see Section \ref{sect53}).
}\end{remark}

\section{Imaging a Gaussian beam}

\subsection{The double conjugation law}

Let ${\cal L}$ be a centred system, afocal or with foci. Let ${\cal A}$ (vertex $V_A$, center of curvature
$C_A$) be a spherical emitter in the object space and let ${\cal A}'$ (vertex $V'$, center of curvature $C'$) be its coherent geometrical image formed by ${\cal L}$. The optical field amplitude $U_{A'}$ on ${\cal A}'$ is a homothetic copy of the field amplitude $U_A$ on ${\cal A}$, that is
\begin{equation}
  U_{A'}(\vec r')={1\over m_{\rm v}}\,U_A\left({\vec r'\over m_{\rm v}}\right)\,,\label{eq23}\end{equation}
where $m_{\rm v}$ is the lateral magnification at vertices. A constant phase factor is omitted in Eq.\ (\ref{eq23}): it represents the phase shift due to the time taken by light to travel from the object to the image (from vertex to vertex). The image is geometrical in the sense that diffraction due to the limited aperture of ${\cal L}$ is not taken into account. By a coherent image we mean an image in which  the phase difference between the vibrations at two arbitrary  points $A'$ and $B'$ of ${\cal A}'$ equals that between the vibrations emitted at the corresponding object points $A$ and $B$ of ${\cal A}$ (the phase is preserved in imaging).

The double conjugation law states that ${\cal A}'$ is the coherent image of ${\cal A}$ if, and only if, $V$ and $V'$ are conjugate (in accordance with paraxial optics) and $C$ and $C'$ are also conjugate (which constitutes a ``coherent'' property of imaging) \cite{PPF1,PPF2,PPF3}.

By applying the double conjugation law, we derive the following results, which will be used in the next section to image certain wavefronts of a Gaussian beam. Let us  assume that ${\cal L}$ is a lens with foci. 
Moreover, let us assume that ${\cal A}$ is centered at the object focus of ${\cal L}$ ($C\equiv F$). Since the conjugate of $F$ is at infinite distance on the optical axis, the spherical cap ${\cal A}'$ is centered at infinity: ${\cal A}'$ is a plane (it is orthogonal to the axis at $V'$, the conjugate of $V$).

Next, let us assume that ${\cal A}$ is a plane orthognal to the axis at $V$, that is, a spherical cap centered at infinity. Then ${\cal A}'$  is a spherical cap whose vertex is $V'$, the conjugate of $V$, and its center of curvature is the image focus $F'$ (because $F'$ is the image of infinity under ${\cal L}$).

\subsection{Location of the waist of the image beam}

\begin{figure}[h]
  \centering
    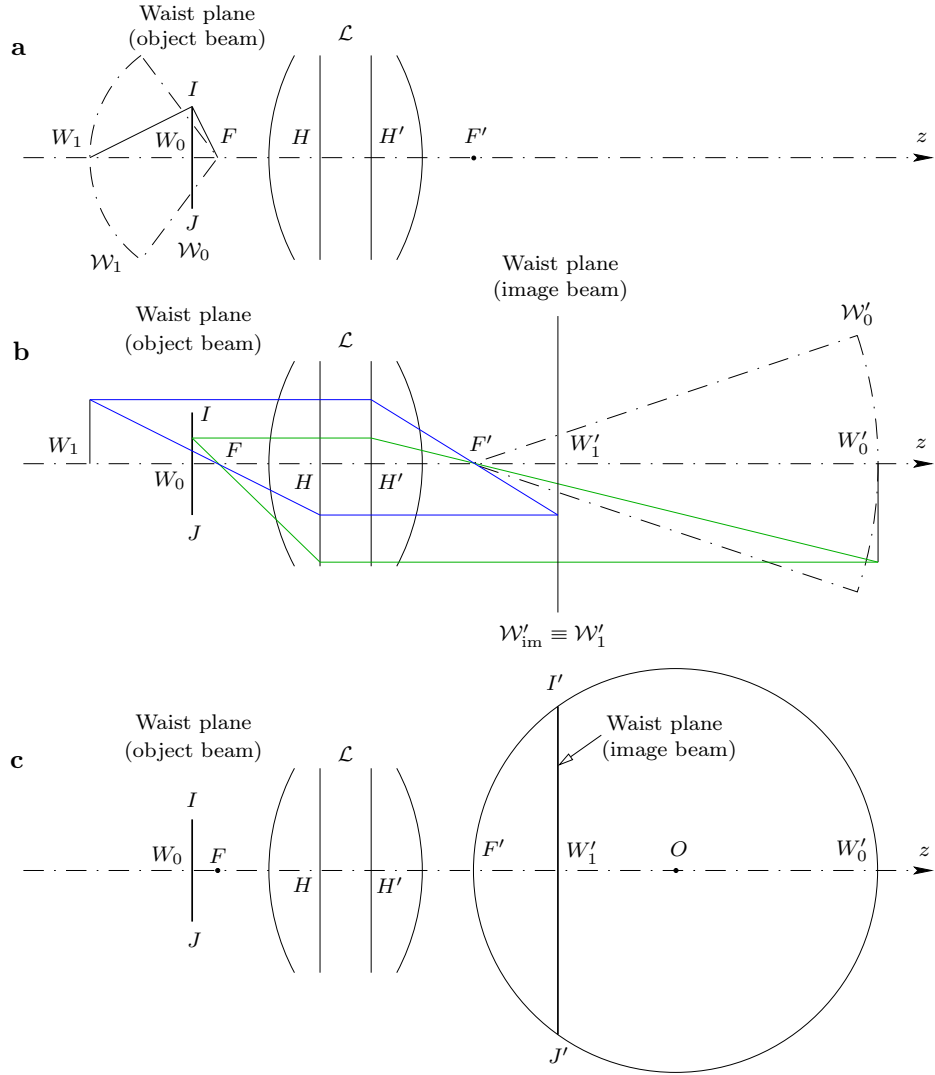
  \caption{\small Construction of the waist plane and the Rayleigh range
    of the image beam formed by a lens (with foci $F$ and $F'$). The point $W'_1$ is the paraxial image of $W_1$, and $W_0'$ is that of $W_0$. The Rayleigh range of the object beam is $\zeta_0=IJ/2$ and that of the image beam is  $\zeta'_0=I'J'/2$. (Points $I'$ and $J'$ are not the images of $I$ and $J$.)\label{fig8}}
\end{figure}

Let us consider a lens ${\cal L}$ whose foci are $F$ and $F'$, and principal points are $H$ and $H'$ (Fig.\ \ref{fig8}). Let ${\cal W}_0$ be the plane of the waist of a Gaussian beam in the object space (located at $W_0$). We recall that the waist of the image beam is not the image of the waist of the object beam  (the reason is that the wavefront at the waist is a plane, and the coherent image of a plane is not a plane, but a spherical cap centered at the image focus of the lens. Imaging a plane gives a plane only for afocal systems). To avoid confusion, we will denote by ${\cal W}'_0$ the wavefront image of the waist plane ${\cal W}_0$ and by ${\cal W}'_{\rm im}$ the waist plane of the image beam.

\goodbreak
 The plane of the waist of the image beam is obtained as follows (see Fig.\ \ref{fig8}):
\begin{enumerate}
\item We start with the plane of the waist (object beam) located at $W_0$, with $W_0I=\zeta_0$. Let $W_1$ be the
  point where the straight line orthognal to $FI$ at $I$ intersects the axis $z$. The spherical cap ${\cal W}_1$ with vertex $W_1$ and center at $F$ is a wavefront of the Gaussian beam (see Section \ref{sect4}).
\item Since ${\cal W}_1$ is centered at the object focus $F$, its image formed by the lens is  the  plane 
  that intersects the $z$--axis  at the point $W'_1$, which is the conjugate of $W_1$, and whose construction is shown in diagram (b) (blue lines). This plane is a wavefront of the image beam, because it is the coherent image of a wavefront in the object space. 
 Since the beam waist of a Gaussian beam is on a plane wavefront, the beam waist of the considered image beam is necessarily located at $W'_1$. Then ${\cal W}'_{\rm im}\equiv {\cal W}'_1$.
\item
  The plane of the waist of the object beam is ${\cal W}_0$, located at $W_0$. Its image is the spherical cap ${\cal W}'_0$ centered at the image focus $F'$ with vertex at $W_0'$, the conjugate of $W_0$. The construction of $W'_0$ is shown in diagram (b) (green lines).
  \item We know a wavefront of the image beam, explicitly, the spherical cap ${\cal W}'_0$ with vertex at $W'_0$ and center at $F'$. To construct the Rayleigh range, we use the construction of Section \ref{sect4}. Let $O$ be the midpoint between $F'$ and $W'_0$ and let ${\cal C}$ be the circle of center $O$ and radius $OF'=OW'_0$. The circle ${\cal C}$ intersects the straight line orthogonal to the axis at $W'_1$, in the points $I'$ and $J'$. The Rayleigh range of the image beam is $\zeta'_0=W'_1I'=I'J'/2$. (We recall that $I'$ and $J'$ are not the conjugates of $I$ and $J$.)
\end{enumerate}

\begin{remark}{\rm  Let $w_1$ be the light spot radius on the wavefront ${\cal W}_1$ (Fig. \ref{fig8} a). Since the waist plane ${\cal W}'_{\rm im}$ of the image beam is the image of ${\cal W}_1$, the transverse radius of the waist of the image beam is $w'_{\rm im}=|m_1|\,w_1$, where $m_1$ is the lateral magnification between point $W_1$ and $W'_1$ (that is $m_1=-f/\overline{FW_1}$, where $f$ is the object focal length of the lens).
  }\end{remark}

\subsection{A particular case}\label{sect53}

We assume that the waist plane of a Gaussian beam is located at the object focus of a lens (foci $F$ and $F'$; focal lengths $f$ and $f'$). We refer to Fig.\ \ref{fig8}. If $W_0$ tends to the object focus $F$, then $W_1$ tends to infinity, so that $W'_1$ tends to the image focus $F'$. We conclude that if the waist of the object beam lies in the object focal plane of the lens, the waist of the image beam lies in the image focal plane. This constitutes a well-known property of Gaussian beam imaging.

\begin{figure}[h]
  \centering
  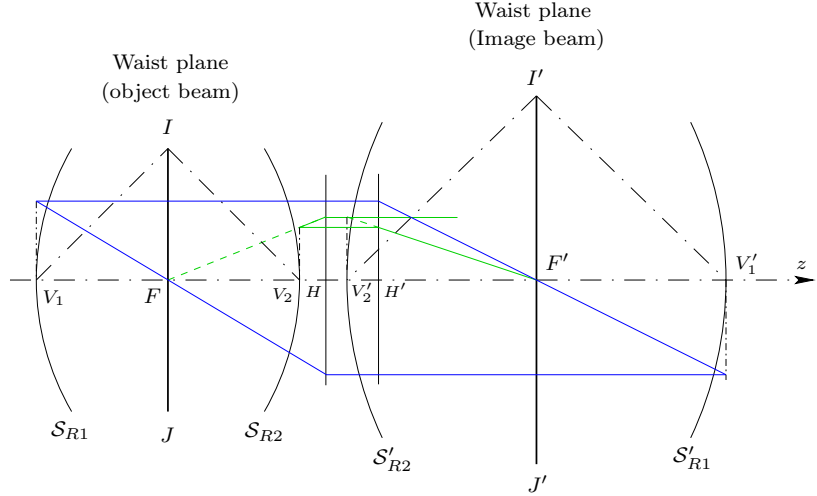
  \caption{\small Construction of the Rayleigh range
    of the image  beam formed by a lens when the waist of the object beam is located at the object focus. The Rayleigh range of the object beam is $\zeta_0=IJ/2=V_1V_2/2$. The lens is represented by its foci $F$ and $F'$, and its principal points $H$ and $H'$ on the axis. The waist of the image beam is located at the image focus of the lens and the corresponding Rayleigh range is $\zeta'_0=I'J'/2=V'_1V'_2/2$. (Points $I'$ and $J'$ are not the images of $I$ and $J$.)\label{fig9}}
  \end{figure}

For an actual contruction of the Rayleigh range of the image beam,
the previous construction fails (because useful spheres lie at infinity). We propose a solution based on geometrical optics. We refer to Fig.\ \ref{fig9}. Let ${\cal S}_{R1}$ (vertex $V_1$) and ${\cal S}_{R2}$ (vertex $V_2$) be the Rayleigh spheres of the object beam. Thus $V_2$ is the center of curvature of ${\cal S}_{R1}$, $V_1$ that of ${\cal S}_{R2}$, and $V_1F=FV_2=IF$.  Let $V'_1$ be the conjugate of $V_1$, and $V'_2$ the conjugate of $V_2$. By the double conjugation law, we obtain that (a) the image  ${\cal S}'_{R1}$, which is a wavefront of the image beam, has its vertex at $V'_1$ and its center at $V'_2$; (b)  the image  ${\cal S}'_{R2}$ has its vertex at $V'_2$ and its center at $V'_1$. That means that ${\cal S}'_{R1}$ and ${\cal S}'_{R2}$ are confocal spheres, because $R'_1=\overline{V'_1V'_2}=-R'_2$. 
The Newton conjugation formula gives $\overline{FV_1}\cdot\overline{F'V'_1}=ff'=\overline{FV_2}\cdot\overline{F'V'_2}$. Since $W_0$ is the midpoint of segment $V_1V_2$ and since $W_0\equiv F$, we have $\overline{FV_1}=-\overline{FV_2}$, so that $\overline{F'V'_1}=-\overline{F'V'_2}$, and $F'$ is the midpoint of $V'_1V'_2$. Since the waist of the image beam is at $F'$, we conclude that ${\cal S}'_{R1}$ and ${\cal S}'_{R2}$ are the Rayleigh spheres of the image beam. Then the Rayleigh range of the image beam is $F'I'=F'V'_2=F'V'_1$.

The construction is given in Fig.\ \ref{fig9}, from the prior knowledge of the Rayleigh range $\zeta_0$ of the object beam. In Fig.\ \ref{fig9}, the lens is represented by its principal planes and principal points $H$ and $H'$ on the axis. The drawing steps are the following:
\begin{enumerate}
\item Vertices $V_1$ and $V_2$ are drawn so that  $FV_1=FV_2=FI=\zeta_0$.
\item Conjugate points $V'_1$ and $V'_2$ are constructed as usual in paraxial optics. The construction of $V'_1$ is given by the  blue lines, and the construction of $V'_2$ by the green lines. (We observe that $V'_2$ and ${\cal S}'_{R2}$ are virtual, in the case considered here).
  \item The Rayleigh range of the image beam is $\zeta'_0=F'I'=F'V'_1=F'V'_2$.
  \end{enumerate}

Eventually, we point out that since $\zeta_0=FV_1$ and $\zeta'_0=F'V'_1$, the Newton conjugation formula $\overline{FV_1}\cdot\overline{F'V'_1}=ff'$ gives $\zeta_0\,\zeta'_0=-ff'$. For the waist radii, we obtain
  \begin{equation}
    w_0^{\; 2}\,{w'_0}^{2}=-{\lambda \lambda ' ff'\over \pi^2}\,.
  \end{equation}
  where $\lambda$ is the wavelength in the object space, and $\lambda '$ in the image space. From $\lambda f=-\lambda ' f'$, we obtain $ w_0^{\; 2}\,{w'_0}^{2}={{\lambda '}^2 {f'}^2/ \pi^2}$,
  and then
  \begin{equation}
    w'_0={\lambda '|f'|\over \pi w_0}\,.\end{equation}

\section{Complement: Waist planes for limiting cases of stable \\ resonators}\label{sect6}

In Section \ref{sect2} we mentioned that an optical resonator formed by two mirrors is stable if the vertices and centers of curvature of the mirrors  are arranged conveniently. If this condition is satisfied, the construction in Section \ref{sect3} provides the waist plane and the Rayleigh range of the resonator.

We now examine what happens for a confocal resonator formed by two spherical mirrors (with radii of curvature $R_1$ and $R_2$ and distance $D$ between the mirrors).

We first show that a confocal resonator is either unstable or constitutes  a limiting case of stability.
Let $V_j$ and $C_j$ ($j=1,2$) be the vertices and the centers of curvature of the mirrors. We assume $V_1\ne V_2$ (that is $D\ne 0$). Since the resonator is confocal, the midpoints between $V_1$ and $C_1$ on one side and between $V_2$ and $C_2$ on the other side coincide. The arrangements of points $V_j$ and $C_j$ are therefore as follows:
\begin{itemize}
\item  If $|R_1|\ne |R_2|$, the  subscripts of $V_j$ and $C_j$ are arranged according to 1221 or 2112, and the resonator is unstable.
\item  If $R_1=-R_2>0$, then $D=R_1$, that is $V_1\equiv C_2$ and $V_2\equiv C_1$.  The resonator is symmetric and confocal, and constitutes a limiting case of stability. (The case $R_1=-R_2<0$ can be treated by permuting the mirrors.) 
  \end{itemize}

In the following, we examine the case of a symmetric confocal resonator. The  circle method (see Section \ref{sect3}) fails to provide the waist plane and the Rayleigh range, because circles ${\cal C}_1$ and ${\cal C}_2$ coincide, so that their intersection is not reduced to two points. 
Actually we will confirm this property and show that the waist plane may be located at any point between the mirrors.

We adapt Fig. \ref{fig1} to a resonator:  the wavefronts ${\cal S}_1$ and ${\cal S}_2$ become the mirrors  ${\cal M}_1$ and ${\cal M}_2$. Light propagates from left to right.  We consider a symmetric confocal resonator, denoted as ${\cal R}$,  with $D=R_1=-R_2>0$.

We will consider various stable resonators ${\cal R}'_\varepsilon$---for which the waist planes are given by the circle method---that tend to  ${\cal R}$,  and we will show that the limit waist plane is not always the same.
The radii of curvature of the mirrors forming the resonator ${\cal R}'_\varepsilon$ are $R'_1$ and $R'_2$ and the algebraic distance from ${\cal M}_1$ to ${\cal M}_2$ is $D'$.

For every  resonator ${\cal R}'_\varepsilon$, we set $R'_2=R_2 <0$.
The algebraic distance $d_1$ from the waist plane of ${\cal R}'_\varepsilon$ to mirror ${\cal M}_1$ is $d_1$ given by Eq.\ (\ref{eq7}), i.e.
\begin{equation}
  d_1=-{D'(D'+R_2)\over 2D'-R'_1+R_2}\,.\label{eq29}\end{equation}

\begin{figure}[h]
  \centering
  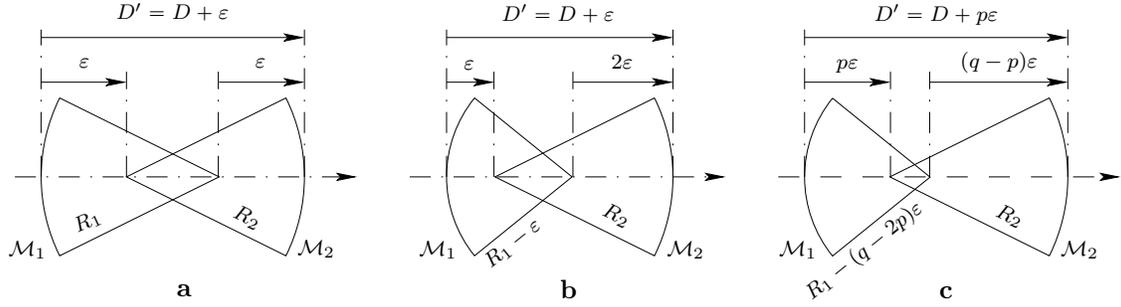
  \caption{\small Three examples of stable resonators that tend to a symmetric confocal resonator when $\varepsilon$ tends to $0$. The confocal resonator is such that $D=R_1=-R_2>0$. The distances from mirror ${\cal M}_1$ to the waist plane are (a) $D'/2$, (b) $D'/3$; and (c) $pD'/q$; in the limit these become (a) $D/2$, (b) $D/3$, and (c) $pD/q$. \label{fig10}}
  \end{figure}

We consider three family of resonators, indexed by the parameter $\varepsilon$ (Fig.\ \ref{fig10}):
\begin{enumerate}
  \item[a.]
Let us consider first a resonator ${\cal R}'_\varepsilon$ with  $R'_1=R_1=-R_2$ and $D'=D+\varepsilon$, $R_1>\varepsilon >0$ (Fig.\ \ref{fig10}--a).   The resonator is stable and Eq.\ (\ref{eq29}) gives
\begin{equation}
  d_1=-{D'(D'+R_2)\over 2D'-R'_1+R_2}=-{D'(D'+R_2)\over 2D'+2R_2}= -{D'\over 2}\,.
\end{equation}
If we move ${\cal M}_1$ to the right (that is, if $\varepsilon$ tends to $0$), $D'$ tends to $D$, so that in the limit $D=R_1$. The resonator ${\cal R}'_\varepsilon$ tends to the initial  symmetric confocal resonator ${\cal R}$, and $d_1$ tends to $-D/2$. The waist plane tends to the midpoint between $V_1$ and $V_2$.

\item[b.] Let us assume $R'_1+\varepsilon=-R_2 =D'-\varepsilon$  with $-R_2>2\varepsilon >0$ (Fig.\ \ref{fig10}--b). The resonator ${\cal R}'_\varepsilon$ is stable.  Then
  \begin{equation}
    d_1=-{D'\varepsilon\over 2(-R_2+\varepsilon) +R_2+\varepsilon+R_2}=-{D'\over 3}\,.
  \end{equation}
  If $\varepsilon$ tends to $0$, $D'$ tends to $D$ and $R'_1$ tends to $R_1$. The resonator ${\cal R}'_\varepsilon$ tends to ${\cal R}$ and, in the limit, its waist plane is at a distance $D/3$ from ${\cal M}_1$.

\item[c.]   The previous exemples can be generalized as follows (Fig.\ \ref{fig10}--c). Let $R'_1$ and $D'$ be such that  $R'_1+(q-2p)\varepsilon=-R_2 =D'-p\varepsilon$, with $q>p >0$ and $-R_2>(q-p)\varepsilon >0$ ($p$ and $q$ may be positive integers). The resonator ${\cal R}'_\varepsilon$ is stable and
   \begin{equation}
    d_1=-{D'p\varepsilon\over 2(-R_2+ p\varepsilon)+R_2+(q-2p)\varepsilon +R_2}=-{pD'\over q}\,.
   \end{equation}
   For every value of $\varepsilon$ we obtain a  resonator ${\cal R}'_\varepsilon$  whose waist plane is located at a distance $pD'/q$ from ${\cal M}_1$. When $\varepsilon$ tends to $0$, we obtain a sequence of stable resonators with  waist planes at a distance $pD'/q$ from ${\cal S}_1$. In the limit, since $D'$ tends to $D$ and $R'_1$ to $R_1$, the sequence of ${\cal R}'_\varepsilon$ tends to ${\cal R}$ whose waist is then at a distance $pD/q$ from mirror ${\cal M}_1$.
\end{enumerate}

We conclude that the waist plane of a symmetric confocal resonator my be located at any point between the mirrors forming the resonator.

\section{Conclusion}

Graphical methods are useful for rapid evaluation of certain properties of Gaussian beams and optical resonators. For example, analyzing the arrangement of vertices and centers of curvature of mirrors along the optical axis immediately indicates whether the resonator is stable.  The graphical constructions presented in this article may help with the quick analysis of Gaussian beams.

Finally, we point out that these constructions may be applied to Gaussian beams with elliptical waists. In such a  case, the previous analysis and drawings are applied twice, independently, in two orthogonal principal cross-sections of the beam.

{\small 
    

}



\end{document}